# Suspended core subwavelength fibers: practical designs for the low-loss terahertz guidance

Mathieu Rozé*[1], Bora Ung*[1], Anna Mazhorova[1], Markus Walther[2] and Maksim Skorobogatiy[1]

* these authors contributed equally to the paper

[1]*Department of Engineering Physics, Ecole Polytechnique de Montréal, C.P 6079, succ. Centre-Ville, Montreal, Quebec, Canada, H3C 3A7*
*www.photonics.phys.polymtl.ca*

[2]*Freiburg Materials Research Center, University of Freiburg, Stefan-Meier-Strasse 21, D-79104, Freiburg, Germany*

**Abstract:** In this work we report two designs of subwavelength fibers packaged for practical terahertz wave guiding. We describe fabrication, modeling and characterization of microstructured polymer fibers featuring a subwavelength-size core suspended in the middle of a large porous outer cladding. This design allows convenient handling of the subwavelength fibers without distorting their modal profile. Additionally, the air-tight porous cladding serves as a natural enclosure for the fiber core, thus avoiding the need for a bulky external enclosure for humidity-purged atmosphere. Fibers of 5 mm and 3 mm in outer diameters with a 150 μm suspended solid core and a 900 μm suspended porous core respectively, were obtained by utilizing a combination of drilling and stacking techniques. Characterization of the fiber optical properties and the near-field imaging of the guided modes were performed using a terahertz near-field microscopy setup. Near-field imaging of the modal profiles at the fiber output confirmed the effectively single-mode behavior of such waveguides. The suspended core fibers exhibit broadband transmission from 0.10 THz to 0.27 THz (larger core), and from 0.25 THz to 0.51 THz (smaller core). Due to the large fraction of power that is guided in the holey cladding, fiber propagation losses as low as 0.02 $cm^{-1}$ are demonstrated. Low-loss guidance combined with the core isolated from environmental perturbations make these all-dielectric fibers suitable for practical terahertz imaging and sensing applications.

## 1. Introduction

In the past few years, interactions between matter and terahertz waves have stimulated research especially for biomedical sensing, noninvasive imaging, non-destructive testing and spectroscopy applications [1-3]. However, because of high material losses in the terahertz range, the design and fabrication of low-loss waveguides for broadband terahertz transmission remains challenging. Several designs of low-loss waveguides have recently been investigated including: metallic wires [4-7], all-dielectric subwavelength polymer fibers [8-11], Bragg fibers [12, 13], and dielectric metal-coated tubes [14-17] to name a few.

A definite advantage of subwavelength dielectric fibers is the ease and highly efficient coupling from a conventional linearly polarized Gaussian-like beam emitted from a terahertz dipole antenna. As a main disadvantage of dielectric fibers we cite their relatively small bandwidth, which is limited towards higher frequencies due to onset of high material absorption losses; and limited at low frequencies by scattering losses [18].

Currently, a key inconvenience of subwavelength dielectric fibers stems from the large fraction of power that is guided outside the high-index core. The latter feature results in strong coupling to the surrounding environment hence making subwavelength fibers difficult to manipulate and to support using holders without disrupting the signal. They also require a bulky gas-purged enclosure to minimize the effects of ambient water vapor on the measured THz spectra.

Recently, a new approach based on introducing porosity in the core, for low-loss terahertz guiding has been proposed by our group [19, 20] and later in [21], and various designs of porous microstructured fibers have been proposed and fabricated [22, 23]. It was also theoretically and experimentally demonstrated that the introduction of porosity enables broadening of the main transmission window compared to a non-porous fiber of the same diameter, and also blue-shifting of the transmission peak to higher frequencies [18].

In this work, we present the analysis of two suspended core all-dielectric fibers specifically designed for practical applications in terahertz guiding. The proposed fiber design incorporates a subwavelength-diameter core suspended inside a large porous outer cladding. We show that the purpose of the porous cladding is two-fold. First, it effectively isolates the core-guided mode from interacting with the surrounding environment, thus preventing undesirable external perturbations to affect the terahertz signal waveform. Second, it serves as a natural air-tight enclosure for the fiber core, thus avoiding the need for an externally purged housing.

The paper is organized as follows: Section 2 describes the geometry and fabrication procedure of the two suspended core fibers. Section 3 presents a detailed analysis and comparison of the output mode profiles yielded by the THz near-field imaging and numerical simulations. Section 4 presents the transmission and propagation losses in both bulk material and fibers inside the 0.01-1.00 THz range, as well as their theoretical modeling. Finally, we discuss the unique properties and potential applications of these fibers.

## 2. Preform and fiber fabrication

All fibers in this work were fabricated using commercial rods of low density polyethylene (PE) known to be one of the lowest loss polymer in the terahertz region [24]. The 12 cm long preform of the solid suspended core fiber was obtained by drilling three holes of 4 mm diameter, equidistantly spaced by 2 mm, and centered on a one-inch diameter rod. The preform was then drawn under pressure into a fiber of 5.1 mm outside diameter and a suspended core of $d_{core}$ =150 µm in size. The whole cross-section of the solid suspended core fiber is presented in Fig. 1(a), with an enlarged view of the core region in Fig. 1(b).

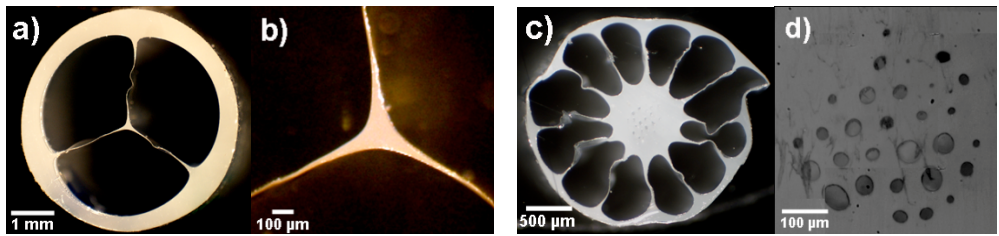

Fig. 1. (a) Cross-section of the suspended core fiber (OD = 5.1 mm), and (b) close-up view of the suspended core region ($d_{core}$ = 150 µm). (c) Cross-section of the porous core fiber (OD = 3 mm), and (d) close-up view of the suspended large porous core ($d_{core}$ = 900 µm)

Another fiber type that we investigated was a suspended large porous core fiber fabricated using a combination of drilling and stacking techniques. Capillaries of 0.8 mm inside diameter (ID) and 1.2 mm outside diameter (OD) were first drawn under pressure from initial PE tubes of 10 mm ID and 25.4 mm OD. Resulting capillaries were then stacked into a hexagonal lattice of 3 rings and solidified in a furnace. The stack of capillaries was then inserted in the middle of a large PE tube presenting 12 holes of 3 mm diameter in its periphery. After pressure-controlled drawing, we obtained a fiber of 3 mm OD with a porous core of approximately 900 µm in diameter suspended in the middle of the holey cladding. The structure of the porous core shows holes ranging from 20 µm to 70 µm in size, resulting in a porosity of approximately 4%. The whole cross-section of the fiber is presented in Fig. 1(c) and a detailed view of the microstructured porous core in Fig. 1(d).

## 3. Modal properties of the waveguides

### 3.1 Near-field characterization of the fiber output

In this section we investigate the principal guiding mechanisms of these fibers. Specifically, we expect that guidance of these fibers is a combination of single-mode guidance inside the subwavelength core, and anti-resonant guiding by the tube of finite thickness. Therefore the main task of this Section is to find out what excitation regimes lead to one or the other regime.

To accomplish this task we use a direct near-field imaging at the fibers output facet to determine unambiguously the nature of the modal composition in these fibers. Near-field images of the output field profiles in both fibers were obtained using a terahertz near-field microscopy system based on the implementation of photoconductive antennae acting first as the coherent *x*-polarized THz pulse source and as a near-field probe. Details of the experimental procedure and setup can be found in [25-27]. This technique allows recording of the temporal evolution of the electric field in close proximity (<30 μm) to the fiber sample illuminated by a broadband THz pulse, with sub-picosecond precision and subwavelength spatial resolution ($\sim \lambda/20$). After Fourier transform of the time-domain data, frequency-dependent near-field images of the fibers in the 0.01-1.00 THz range were retrieved. Taking the *z*-axis as the direction of propagation in the waveguide, two-dimensional profile maps (*x-y* distribution) of the transverse Ex-field were obtained via raster scanning of the fiber cross-section by the probe detector (oriented along the *x*-direction), yielding a 60 × 60 pixels resolution in a 6 mm$^2$ area that covered the whole output facet of fibers.

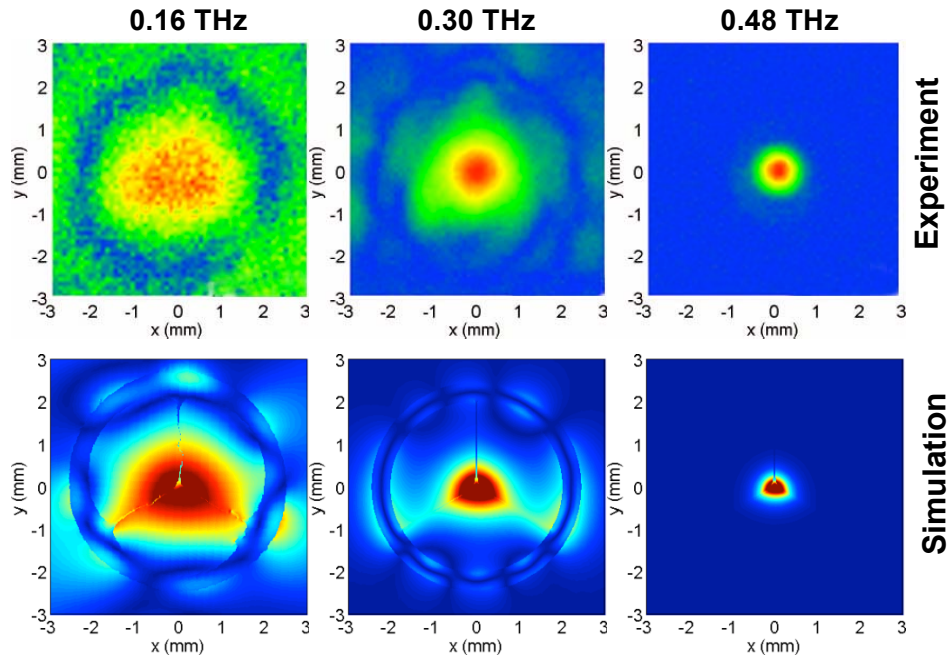

Fig. 2  Near-field microscopy images (upper row and Media 1) and corresponding simulations (lower row) of the output Ex-field profile of the suspended small solid core fiber at 0.16, 0.30 and 0.48 THz.

To understand the principal guiding mechanisms in these fibers, one has to recognize a somewhat complex modal structure in these waveguides. Thus, even if propagation in the suspended core is typically single-mode, the fiber can also support a variety of cladding modes. Then, if the excitation beam is large enough (which is definitely the case at low frequencies) one should expect contributions from both core modes and cladding modes in the total transmission of the fiber.

In Fig. 2 and Media 1 we present |Ex|-field output profiles at selected frequencies as measured experimentally (top row) and as computed numerically (bottom row) for the case of a small suspended-core fiber. From these figures one can distinguish two regimes of propagation. The first regime is anti-resonant guidance at 0.16 THz where the fiber acts as a capillary tube also known as ARROW fibers (modal properties of THz ARROW fibers have recently been discussed in detail in [28]). In that regime, the modal field is strongly delocalized and extends far away from the suspended core, and the guiding mechanism is essentially dictated by the Fabry-Pérot resonant conditions in the tube cladding of finite thickness. One notices in Fig. 6(a) a narrow transmission peak at 0.16 THz corresponding to resonant field confinement in Fig. 2 at that frequency.

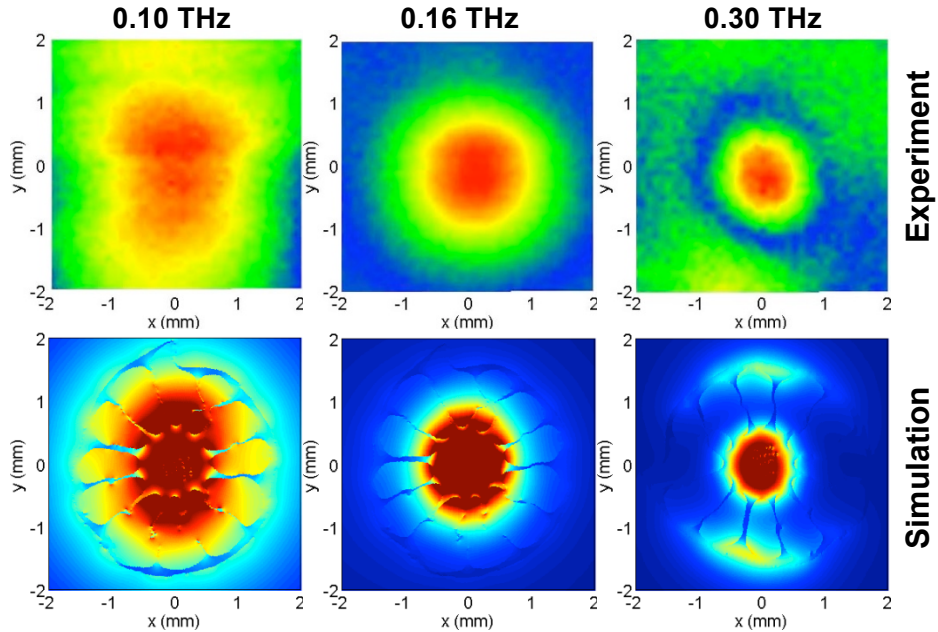

Fig. 3  Near-field microscopy images (upper row and Media 2) and corresponding simulations (lower row) of the output Ex-field profile of the suspended large porous core fiber at 0.10, 0.16 and 0.30 THz.

The second propagation regime in Fig. 2 is the main regime of interest for us where the field is confined in the central solid core and guided by total internal reflection as illustrated by the output profiles at 0.30 THz and 0.48 THz. One also notes that these guided modes are located inside the main low-loss propagation window (as given by the cut-back measurements) [Fig. 7(a)] spanning the range between 0.28 and 0.48 THz. Moreover, the near-field profiles [in Fig. 2] indicate that transmission occurs in an effectively-single mode regime. Inspection of the values for the coupling coefficients of the first $N=12$ modes (not shown here for brevity) confirms that the $HE_{11}$ fundamental mode is dominantly excited inside this frequency range. As expected [in Fig. 2], field confinement becomes stronger as the frequency increases such that for $f > 0.50$ THz practically all the power propagates within the lossy solid core, thus explaining the steep increases of propagation losses over the level of bulk losses thereafter [see Fig. 7(a)].

Figure 3 shows |Ex|-field distributions at the output facet of a large suspended-porous-core fiber at selected frequencies. One observes a similar behavior, as previously described for the small suspended-core fiber, albeit shifted towards lower frequencies. In particular, at low frequencies (close to 0.10 THz) the field is highly delocalized and extends in the microstructured holey cladding, thus enabling attenuation values smaller than the bulk material loss [Fig. 7(b)] due to the large fraction of power guided in low-loss air. For frequencies larger than 0.16 THz, most of the power propagates within the large porous core thus leading to increase in the attenuation loss due to material absorption. Modal confinement in the fiber core becomes stronger at higher frequencies, thus explaining a rapid decline in transmission above 0.20 THz [see Fig. 6(b)].

*3.2 Details of the numerical modeling of the fields at the fiber output*

The distribution of the transverse E-field components $\vec{E}_{output} = \left( E^x_{output}, E^y_{output} \right)$ at the output facet of a waveguide of length $L_w$ is modeled as the coherent superposition of $N$ guided modes (including both core and cladding modes):

$$\vec{E}_{output}(x,y,\omega) = \sum_{m=1}^{N} C_m \cdot \vec{E}_m(x,y,\omega) \cdot e^{i\frac{\omega}{c}\left(n_{eff,m} L_w\right)} e^{-\frac{\alpha_m L_w}{2}} \quad (1)$$

where $\vec{E}_m = \left( E^x_m, E^y_m \right)$ stands for the transverse field components of the *m*-th guided mode. The variables $\alpha_m$ and $n_{eff,m}$ denote respectively the power loss coefficient and the real effective index of the *m*-th mode at a given frequency $\omega$. The variable $C_m$ refers to the normalized amplitude coupling coefficients computed from the overlap integral of the respective flux distributions of the *m*-th mode with that of the input Gaussian beam. Specifically, the definition of $C_m$ is based on the continuity of the transverse field components across the input interface (i.e. cross-section of the subwavelength fiber) between the incident Gaussian beam and the excited fiber modes:

$$C_m = \frac{1}{4} \int \left[ E^{x*}_{input}(x,y) \cdot H^y_m(x,y) + E^x_m(x,y) \cdot H^{y*}_{input}(x,y) \right] dx\, dy \quad (2)$$

where the electromagnetic fields of each mode are normalized to carry unit power in the $z$-direction $\vec{F}/\sqrt{\frac{1}{2}\int \text{Re}(\vec{E}_t \times \vec{H}_t^*)dx\,dy}$ where $\vec{F}$ denotes electromagnetic field vector. Since the probe antenna is placed directly at the fiber output facet, use of the output coupling coefficients is not required. To model the input source, we assume an $x$-polarized Gaussian beam whose fields are fields normalized to carry power $P$:

$$\vec{E}_{input}(x,y) \simeq \hat{x} \cdot \sqrt{\frac{2P}{\pi\sigma^2 n_{clad}}} \cdot \exp\left[-\frac{(x^2+y^2)}{2\sigma^2}\right]$$

$$\vec{H}_{input}(x,y) \simeq \hat{y} \cdot \sqrt{\frac{2P \cdot n_{clad}}{\pi\sigma^2}} \cdot \exp\left[-\frac{(x^2+y^2)}{2\sigma^2}\right] \quad (3)$$

where the Gaussian beam waist parameter ($\sigma$) is related to the full-width half-maxima through $\text{FWHM} = 2\sigma\sqrt{2 \cdot \ln 2}$ and $n_{clad}$ is the refractive index of the cladding medium. The frequency dependence of the beam waist was measured independently by the same near-field microscopy setup [see Fig. 4], and then fitted by a linear function of the input wavelength $\sigma \approx (0.894)\lambda$. This model was subsequently used in the following simulations.

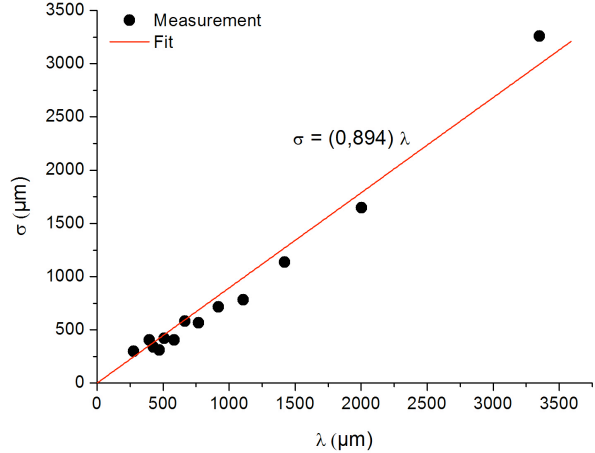

Fig. 4  Gaussian beam waist parameter [$\sigma$ in Eq. (5)] as a function of input wavelength as measured by THz near-field microscopy (dots), and modeled by a linear fit (solid line).

Note that Eq. (1) also requires the knowledge of the frequency dependent material loss of the fiber. This data was measured experimentally, and details are presented in the next section.

Finally, a finite-element method (FEM) code was used to calculate the modes of the fibers. To perform such simulation, we first imported the fiber cross-section geometries (as captured by the optical microscope) into

COMSOL Multiphysics FEM software, and then solved for the complex effective refractive indices and field profiles of the first $N$ modes (both core guided and cladding modes) where $N=12$ for the suspended small solid core fiber, and $N=8$ for the suspended large porous core fiber.

## 4. Fiber transmission and material loss measurements

### 4.1 Bulk polymer material: refractive index and absorption losses

We first report on the refractive index and absorption losses of the commercial polyethylene bulk material used in fiber fabrication. The data is presented in Fig. 5(a) and Fig. 5(b) respectively. Characterization of refractive index and absorption losses was performed with a THz-TDS setup using thick polymer slabs with parallel interfaces. The sample was prepared by cutting and polishing a 1.5 cm thick slice of the rod used for the fiber preform.

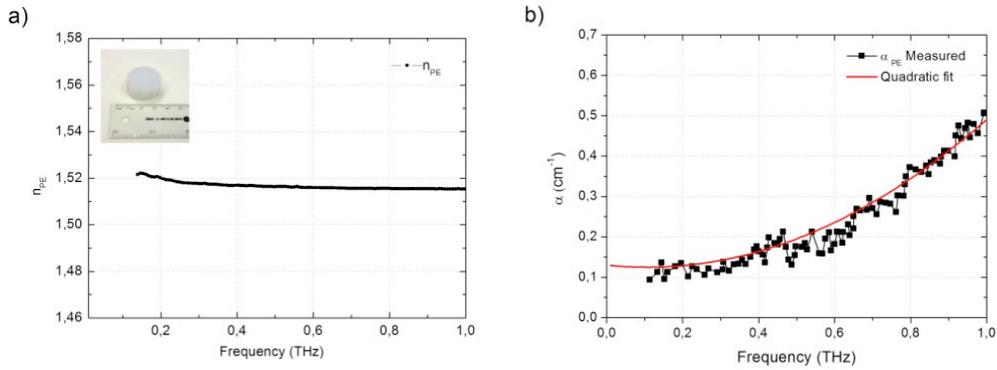

Fig. 5. Refractive index (a) and losses (b) of polyethylene between 0.10 THz and 1THz. The inset picture in Fig. 2(a) presents the polyethylene slab used for both measurements.

The refractive index and absorption losses of polyethylene were retrieved by fitting the predictions of a transfer matrix model to the experimental transmission data [29-30]. The fitted refractive index [Fig. 5(a)] is largely constant between 0.10-1.00 THz and equal to $n_{PE} = 1.514$. Power absorption losses in $cm^{-1}$ of polyethylene increases quadratically as a function of frequency and can be fitted as: $\alpha\,[cm^{-1}] = 0.46\,f^{2} - 0.1\,f + 0.13$ where $f$ is the frequency in (THz) [solid line in Fig. 5(b)]. The absorption losses reach 0.2 $cm^{-1}$ at 0.5 THz and about 0.5 $cm^{-1}$ at 1 THz.

### 4.2 Fiber transmission and loss measurements

Next, transmission characteristics of both fibers [see Fig. 6] were measured using the same THz-TDS imaging setup described in Section 3; except this time the near-field probe was not scanned over the whole output facet of the fiber. Instead, the probe remained positioned at a single spot located in the center $(x_0, y_0)$ of the fiber core. We would like to note that this approach for measuring fiber transmission loss is somewhat different from a traditional one that measures the total power coming out of the whole fiber cross-section.

Particularly, by using a near field probe placed in the fiber center during cutback measurements one preferentially measures losses of the few lowest order modes. This is related to the fact that such modes have their intensity maxima in the vicinity of the fiber core, thus providing the dominant contribution to the total transmission. In the case of a few mode fiber, this method preferentially measures the loss of a fundamental $HE_{11}$ mode.

Using Eq. (1) we can now derive expression for the intensity of a transmitted field as measured by the near field detector:

$$T_{fiber}(\omega) \approx \left|\vec{E}_{output}(x_0, y_0, \omega)\right| = \left|\sum_{m=1}^{N} C_m \cdot \vec{E}_m(x_0, y_0, \omega) \cdot e^{i\frac{\omega}{c}\left(n_{eff,m} L_w\right)} e^{-\frac{\alpha_m L_w}{2}}\right| \quad (4)$$

where $(x_0, y_0)$ denotes the coordinates of the fiber cross-section center.

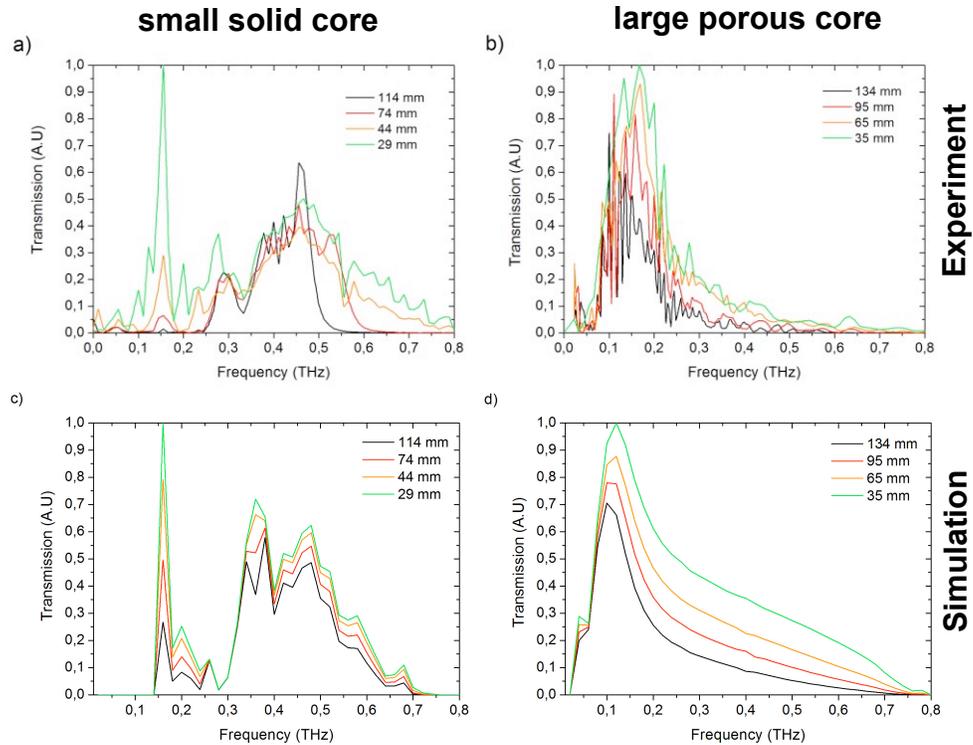

Fig. 6  Ex-field amplitude as measured by the near-field THz-TDS probe between 0.01 and 0.80 THz for the case of (a) suspended small core fiber, (b) suspended large porous core fiber. Corresponding numerical simulations of the amplitude transmission through (c) suspended small solid core fiber, (d) suspended large porous core fiber.

In Figures 6(a) and 6(b) we present experimentally measured transmission spectra for different lengths of the suspended small core fiber and suspended large porous core fiber. The corresponding simulated transmission spectra of both fibers, as modeled by Eq. (4), are presented respectively in Fig. 6(c) and Fig. 6(d).

We note that experimentally measured transmission spectra are well explained by the numerical simulations. Transmission through a suspended small solid core fiber is relatively broadband, covering the 0.25-0.51 THz region. A notable feature in the measured transmission spectrum of the suspended small solid core fiber [Fig. 6(a)] is presented by the sharp transmission peak located at 0.16 THz. This peak is attributed to the coupling of the Gaussian excitation beam to the cladding and surface modes that are plentiful at low frequencies. This can be also confirmed directly by looking at the near-field image of the output mode profile in Fig. 2 at 0.16 THz. Note that cladding modes exhibit high losses due to their strong confinement inside the thick and lossy cladding region. Surface modes also exhibit high losses due to their evanescent nature.

In the case of the suspended large porous core fiber, we first note a transmission window that is narrower [Fig. 6(b)] compared to the small solid core fiber. Specifically for the longest fiber $L_w$=134 mm, frequencies between 0.10 THz and 0.27 THz are effectively guided by such a fiber with a peak transmission near 0.13 THz, which is also reproduced by the numerical modeling [Fig. 6(d)]. One also notices that the transmission abruptly drops at frequencies higher than 0.33 THz, corresponding to a wavelength of 900 μm, which matches the diameter of the fiber porous core. The sharp transmission drop at higher frequencies can be explained by strong confinement of the guided modes inside the lossy fiber core. This rationale is also supported by the near-field images of the guided modes [see Fig. 3 and Media 2] which clearly confirm that for frequencies below 0.30 THz the field is significantly delocalized and extends into the low-loss air cladding; while for frequencies above 0.30 THz the modal fields remains primarily confined inside the porous core. Significant fringe oscillations are visible in the measured transmission spectrum of the suspended large porous core fiber [Fig. 6(b)]. These oscillations are not caused by interference of higher-order modes (or cladding modes) with the fundamental mode. Instead, these closely spaced spectral oscillations represent residual noise stemming from the application of the discrete Fourier transform to retrieve the output spectrum. In fact, simulations [Fig. 6(d)] demonstrate a very smooth transmission spectrum that confirms the effectively single-mode regime in the fiber. The latter claim is supported by the near-field output profiles [Fig. 3] that clearly show the core-guided fundamental mode.

In Figs. 7(a)-(b) we present the power propagation losses of the fibers obtained from cutback measurements (black solid line) and also show (in dashed line) the quadratically increasing bulk material absorption losses obtained from measurements in Fig. 5(b). Upon examining the cutback propagation losses of the small solid core fiber [Fig. 7(a)], we find a minimum value of about 0.02 cm$^{-1}$ inside the 0.28-0.48 THz range, which represents a considerably lower value than the bulk material losses. This is achieved thanks to the large fraction of power guided in the low-loss air cladding, as clearly revealed in the near-field profiles of Fig. 2. At lower frequencies, the highly delocalized field enhances scattering on structural imperfections such as geometrical variations in the bridges thickness. The strongly delocalized mode also enhances the field interaction with the polymer tubular cladding thus inducing high losses inside the 0.01-0.28 THz range. In contrast for frequencies higher than 0.48 THz, the guided power becomes more confined in the suspended small solid core. At these higher frequencies, fiber propagation loss increases dramatically to reach the bulk material loss level, and even surpass it due to scattering losses on microstructural imperfections.

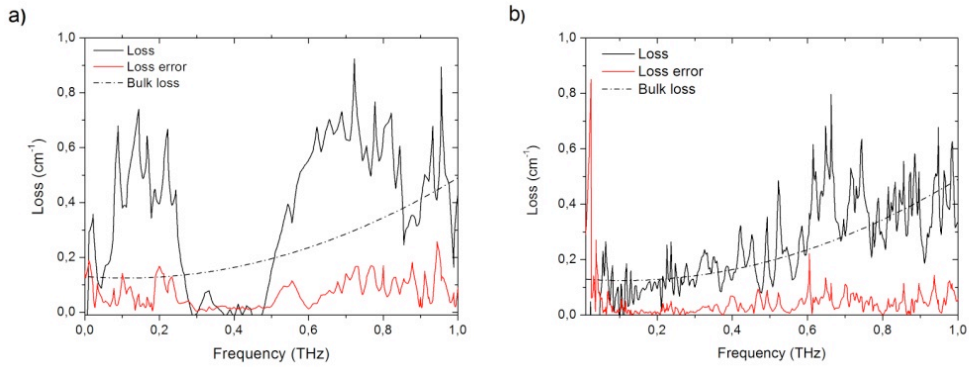

Fig. 7 (a) Power propagation losses, measured by cutback, of the suspended small solid core fiber, and (b) porous core fiber as a function of frequency. Dashed line corresponds to quadratic fit of the bulk material losses.

In Fig. 7(b), corresponding to the suspended large porous core fiber, the general trend of the propagation losses is to follow that of the bulk material loss curve (in dashed line). Propagation losses increase from 0.05 cm$^{-1}$ to 0.15 cm$^{-1}$ between 0.10 THz and 0.40 THz; while bulk material losses quadratically rise from 0.12 cm$^{-1}$ to 0.16 cm$^{-1}$ in the same frequency range. The microstructured holey cladding surrounding the suspended core allows a substantial fraction of power to be guided in air, especially at low frequencies, so that propagation losses are largely lower than the bulk value between 0.05 THz to 0.20 THz: a span which roughly matches the transmission bandwidth identified in the fiber transmission spectrum [Fig. 6(b)]. We note however that beyond 0.20 THz, the 4% porosity of the core is not large enough so as to significantly lower the propagation losses below the level of bulk material absorption. We are presently working on ways to improve the fabrication of the suspended large porous core fiber to significantly increase the core porosity.

## 5. Conclusion

We demonstrate for the first time in this work, the fabrication and near-field characterization of the two suspended core polymer fibers conveniently encapsulated for practical applications. Experimental measurements were confirmed by the full-vector finite-element simulations and analytical modeling. These effectively single mode fibers were designed to support large diameter modes in order to enhance their excitation by the diffraction limited Gaussian beam of a typical THz source. At the same time, the fiber cores are suspended inside of ~3-5 mm-size tubular enclosures filled with dry air to reduce their interaction with the environment, which makes such fibers convenient to handle in practical applications. Two different fiber designs were investigated, one featuring a subwavelength solid core fiber and another one featuring a large porous fiber; in both cases the cores were suspended by the network of ~10 micron-thick bridges inside a much larger diameter tube. In a stark difference with the case of bare subwavelength fibers, the power guided in suspended core fibers is isolated from external disturbances by the tubular dielectric cladding. This feature makes the suspended core fibers excellent candidates for practical terahertz signal delivery for THz near-field imaging and microscopy setups. Moreover, thanks to the highly porous structure, one might envisage the use of suspended core fibers in THz sensing and spectroscopy applications with microfluidics integrated directly into the fiber structure.

The suspended small solid core fiber, in particular, offers very low-loss (0.02 cm$^{-1}$) single mode guiding inside a broad continuous bandwidth (0.28-0.48 THz). This is possible due to a very large aspect ratio (34:1) between the tubular enclosure and core diameters. We note that losses can be further reduced on the low-frequency side by extending the length of the bridges, thus increasing the spacing between the central core and the polymer cladding so as to provide more "room" for the fundamental mode. Moreover, the transmission window can still be further extended on the high-frequency side by reducing the core size and/or by incorporating substantial porosity within the core. The most important advantage provided by these suspended core fibers stems from the tubular cladding which effectively shields the core, and the propagating signal it supports, from perturbations in the surrounding environment therefore allowing convenient hand manipulation and positioning with holders. This last crucial property suggests that suspended core fibers offer a promising route towards practical all-dielectric THz waveguides.